\documentstyle[11pt,twoside,jltp]{article}

\title{Vortices Observed and to be Observed}

\author{G.E. Volovik\address{Low Temperature Laboratory,
Helsinki University of Technology,
P.O.Box 2200, FIN-02015 HUT, Finland\\
L.D. Landau Institute for
Theoretical Physics,  Kosygin Str. 2, 117940 Moscow, Russia}}

\runninghead{G.E. Volovik}{Vortices Observed and to be Observed}

\begin{document}

\begin{abstract}
Linear defects are generic in continuous media. In quantum systems they appear
as topological line defects which are associated with a circulating persistent
current. In relativistic quantum vacuum they are known as cosmic
strings, in superconductors as quantized flux lines, and in superfluids and
low-density atomic Bose-Einstein condensates as quantized vortex lines. We
discuss unconventional vortices in unconventional superfluids and
superconductors, which have been observed or have to be observed, such as
continuous singly and doubly quantized vortices in $^3$He-A and chiral
Bose condensates; half-quantum vortices (Alice strings) in $^3$He-A and in
nonchiral Bose condensates; Abrikosov vortices with fractional magnetic flux in
chiral and $d$-wave superconductors; vortex sheets in $^3$He-A and chiral
superconductors; the nexus -- combined object formed by vortices and
monopoles.
Some properties of vortices related to the fermionic quasiparticles living in
the vortex core are also discussed.

\end{abstract}

\maketitle


\section{INTRODUCTION}

Phases of superfluid $^3$He provided the first example of unconventional
superfluidity, which triggered the search for the identification of the
unconventional superconductivity and superfluidity first in the systems with
heavy fermions, then in high-temperature superconductors, and most recently
in a
layered superconductor Sr$_2$RuO$_4$ and in a multi-component Bose condensates.
Unconventional superfluidity of $^3$He-A manifests itself in exotic types
of the
topological defects and in unusual behavior at low temperature caused by the
gap nodes. $^3$He-A is the first example of superfluids with the gap nodes
in the
energy spectrum for quasiparticles. Though existence of nodes in $^3$He-A
was proven only recently in experiments with the AB interface where the $T^4$
behavior has been verified \cite{Fisher}, these nodes influence the
investigation of low-$T$ properties of unconventional
superconductors.  In particular, the observed low-$T$ and low-$B$ scaling  for
the specific heat in a mixed state of high-$T_c$ superconductors
related to the gap nodes and vortices \cite{Revaz} has its history started in
1981, when it was found that gap nodes in $^3$He-A lead to finite nonanalytic
density of states at $T=0$ in the presence of the order parameter texture
\cite{VolMin1981}. The intermediate steps were the Ref.
\cite{Muzikar1983,Nagai1984}, where the superfluid velocity field
has been used instead of a texture, and Ref. \cite{Volovik1993} where this was
extended for the superflow around the vortex in $d$-wave superconductor.

Another example of influence of $^3$He-A is the observation in
high-$T_c$ superconductors of the fractional flux attached to the
tricrystal line, which is the junction of three grain
boundaries\cite{Kirtley1996}. This was inspired by the discussion of the flux
quantization with the half of conventional magnitude in a heavy fermionic
superconductors \cite{Geshkenbein1987}, which in turn has been inspired by
$^3$He-A where the possibility of a vortex with half-integer circulation has
been shown \cite{VolMin1976}.

The  two
examples above are the two sides of the same phenomenon: the nontrivial
topology. The gap nodes in momentum space and vortices -- the nodes in position
space -- are topological objects, which can be
continuously transformed to each other by transformation in the extended
momentum+position space, as  also was first discussed in $^3$He-A
\cite{VolMin1982}. The same topology is also in the basis of the Standard Model
of electroweak and strong interactions in particle
physics \cite{NielsenNinomiya} (see  Review \cite{Volovik2000}). As a
result, in
the vicinity of the gap nodes the $^3$He-A  liquid as well as the
high-temperature superconductors and other similar systems acquire some
attributes of relativistic quantum field theory such as  Lorentz invariance,
gauge invariance, chirality and chiral fermions, etc.  The conceptual
similarity
between these condensed matter systems and the quantum vacuum makes them an
ideal laboratory for simulating many effects in high energy
physics and cosmology.

Now it is believed that the tetragonal layered superconductor
Sr$_2$RuO$_4$ has the order parameter of the same symmetry class as  $^3$He-A
\cite{Rice,Ishida} and thus shares many unusual properties of the latter. In
particular, it is assumed that the superconducting state of Sr$_2$RuO$_4$
is chiral, i.e.  reflection and time reversal symmetries are
spontaneously broken. There are other systems which
share some of unusual properties of $^3$He-A: the multicomponent atomic
Bose-condensates \cite{Ho,Ohmi}; neutron superfluids; quark matter
with colour superconductivity ...  It is worthwhile to consider where we are
now: what is observed from these exotic properties and what is to be observed.

\section{CONTINUOUS VORTICITY}

According to the  Landau picture of superfluidity,
the superfluid flow is potential: its velocity ${\bf v}_s$ is curl-free:
$\nabla\times {\bf v}_s$. Later Onsager \cite{Onsager} and Feynman
\cite{Feynman} found that this statement must be generalized:
$\nabla\times {\bf v}_s \neq 0$  at singular lines, the quantized
vortices, around which the phase of the order parameter winds by $2 \pi N$.
Discovery of superfluid
$^3$He-A made further weakening of the rule: The nonsingular vorticity can be
produced by the regular texture of the order parameter \cite{Mermin-Ho}.
The order parameter -- the wave function of Cooper pair with spin $S=1$ and
orbital momentum $L=1$ --  is determined by two vectors
\begin{equation}
\Delta({\bf k})=  \Delta_0 ~(\sigma\cdot\hat {\bf d})~\left( {\bf k}\cdot (\hat
{\bf e}_1 + i\hat {\bf e}_2) \right)~.
\label{OrderParameter}
\end{equation}
The real unit vector $\hat{\bf d}$ enters the spin part of the order
parameter, where ${\bf\sigma}$ are the Pauli matrices. The orbital part, which
shows the momentum
${\bf k}$ dependence, is described by the complex vector $\hat{\bf
e}_1+i\hat{\bf e}_2$, whith mutually orthogonal real and imaginary parts. The
order parameter is intrinsically complex, i.e. its phase cannot  be eliminated
by a gauge transformation. This violates the time reversal symmetry and
leads to
the spontaneous angular momentum along the unit vector $\hat{\bf l}=\hat{\bf
e}^{(1)}\times\hat{\bf e}^{(2)}$. The superfluid velocity of the chiral
condensate is ${\bf v}_s={\hbar\over 2m}\hat e^{(1)}_i\nabla \hat e^{(2)}_i$
(where $2m$ is the mass of the Cooper pair) with continuous vorticity
satisfying
the Mermin-Ho relation: $
\nabla\times {\bf  v}_s = {\hbar\over 4m~}e_{ijk} \hat l_i   \nabla \hat
l_j\times \nabla\hat l_k $.

Quantization of vorticity is provided by topology of
textures: The texture which is homogeneous at large distances
\cite{AndersonToulouse}  has $4\pi$ winding (i.e. $N=2$) if the $\hat{\bf l}$
field covers all $4 \pi$ orientations in the soft core of continuous vortex.
If the order parameter is a spinor (the ``half of vector''), as it
occurs in spin-1/2 Bose condensates \cite{HoTalk} and in the Standard Model of
electroweak interactions, the continuous vortices and correspondingly
continuous  cosmic strings \cite{AchucarroVachaspati} have twice less winding
number, $N=1$.

The NMR measurements already in 1983 provided the indication for
existence of the continuous $4\pi$ lines in rotating $^3$He-A (see Review
\cite{EltsovKrusius}). However, the first proof
that the isolated continuous vortex contains $N=2$ circulation quanta has been
obtained only recently \cite{Blaauwgeers}.  Lines with continuous vorticity has
been used for ``cosmological'' experiments. Investigation of the dynamics of
these vortices presented the first demonstration of the condensed matter analog
of the axial anomaly in relativistic field theory, which is believed
responsible
for the present excess of matter over antimatter \cite{Bevan}. Another
cosmological phenomenon is related to nucleation of these vortices
investigated
in \cite{RuutuCritVel}. Vortex formation marks the helical instability of the
normal/superfluid counterflow in $^3$He-A (the counterflow is the flow of the
normal component with respect to the superfluid one). The instability is
described by the same equations and actually represents the same physics (see
Review \cite{Volovik2000}) as the helical instability of the bath of
right-handed
electrons towards formation of the helical hypermagnetic field discussed in
\cite{JoyceShaposhnikov,GiovanniniShaposhnikov}. This experiment thus
supported
the Joyce-Shaposhnikov scenario of formation of  primordial cosmological
magnetic
field.

Another example of continuous vorticity observed in $^3$He-A is the vortex
sheet \cite{Sheet}. This is the chain of the so called Mermin-Ho
continuous vortices with $N=1$ trapped by the soliton. Each vortex
is the kink in the soliton structure. The kink can live only
within the soliton, precisely as the Bloch line within the Bloch wall in
magnets.

\section{FRACTIONAL VORTICES IN SUPERFLUIDS AND SUPERCONDUCTORS}

In $^3$He-A the fractional vorticity is still to be observed. The discrete
symmetry, which supports the half-quantum vortex comes from the identification
of points
$\hat{\bf d}$,  $\hat{\bf e}_1+i\hat{\bf e}_2$ and $-\hat{\bf d}$,
$-(\hat{\bf e}_1+i\hat{\bf
e}_2)$. In a half-quantum vortex both vectors change sign after circling
around the vortex  while
the order parameter in Eq.(\ref{OrderParameter}) returns to its initial value
\begin{equation}
\hat {\bf d}=\hat {\bf x}\cos {\phi\over 2} +\hat {\bf y}\sin {\phi\over
2}~~,~~
 \hat{\bf e}_1+i\hat{\bf e}_2=(\hat {\bf x}+i\hat {\bf y})e^{i\phi/2}~.
\label{HalfQuantumVortex}
\end{equation}
Here $\phi$ is the azimuthal angle of the cylindrical coordinate system; the
magnetic field is applied along $z$ to keep the $\hat{\bf d}$-vector in $x-y$
plane.  $\hat{\bf d}$ is quantization axis for
spin. Since it rotates by $\pi$ around the vortex,  a ``person'' living in
$^3$He-A, who moves around the vortex, insensibly finds its spin
reversed with respect to the fixed environment.  This is an analog of Alice
string in particle physics \cite{Schwarz}. A particle which moves around an
Alice
string continuously flips its charge or parity or enters the ``shadow'' world
\cite{Silagadze}.

In superconductors the crystalline structure must be taken into account. In
simplest representation, which preserves the tetragonal symmetry, the order
parameter in  Sr$_2$RuO$_4$:
\begin{equation}
\Delta({\bf k})=\Delta_0~(\hat{\bf d}\cdot{\bf \sigma})\left(\sin {\bf
k}\cdot{\bf a}  +
i~\sin {\bf k}\cdot{\bf b} \right)e^{i\theta}~,
\label{ChiralOP}
\end{equation}
where $\theta$ is the phase of the order parameter; ${\bf a}$ and
${\bf b}$ are the elementary vectors of the crystal lattice within the layer.

Vortices with fractional $N$ can be obtained in two ways. If  $\hat {\bf
d}$-field is flexible enough, there is analog of $N=1/2$ vortex in
Eq.(\ref{HalfQuantumVortex}), in which $\hat {\bf d}
\rightarrow -\hat {\bf d} $ and $\theta \rightarrow \theta + \pi$ around the
vortex \cite{Kee}. Another is the M\"obius strip geometry produced by
twisting the crystal axes ${\bf a}$ and
${\bf b}$ around the loop  \cite{Monopoles}. The closed wire traps the
fractional flux, if it is twisted by an angle
$\pi/2$ before gluing the ends. Since the local orientation of the crystal
lattice continuously changes by $\pi/2$ around the loop, ${\bf a} \rightarrow
{\bf b} $ and ${\bf b} \rightarrow
-{\bf a}$, the order parameter is multiplied by $i$ after encircling the loop.
The single-valuedness of the order parameter requires that this must
be compensated by a change of its phase $\theta$ by $\pi/2$. As a result the
phase winding around the loop is $\pi/2$, i.e.  $N=1/4$.

This, however, does not mean that the loop  of the chiral $p$-wave
superconductor traps 1/4 of the magnetic flux $\Phi_0$ of conventional
Abrikosov
vortex. Because of the breaking of time reversal symmetry in chiral crystalline
superconductors, persistent electric current arises not only due to the phase
coherence but also due to deformations of the crystal
\cite{VolovikGorkov1984}:
\begin{equation}  {\bf j}=\rho_s\left( {\bf v}_s-{e\over mc}
{\bf A}\right) + Ka_i\nabla b_i~,~{\bf v}_s={\hbar \over 2m}\nabla \theta ~.
\label{SuperfluidCurrent}
\end{equation}
Magnetic flux trapped by the loop is obtained from the condition of
no current in the wire,  ${\bf
j}=0$ in Eq.(\ref{SuperfluidCurrent}). Thus, the trapped flux
depends on the parameter $K$ in the deformation current. In the
limit case of $K=0$ the flux is $\Phi_0/4$ (or $\Phi_0/6$ if the underlying
crystal lattice has hexagonal symmetry).

In the nonchiral $d$-wave superconductor in layered cuprate oxides
the order parameter can be represented by:
\begin{equation}
\Delta({\bf k})=\Delta_0~\left(\sin ^2{\bf k}\cdot{\bf a} -\sin^2{\bf
k}\cdot{\bf b} \right)e^{i\theta}~.
\label{DWaveOP}
\end{equation}
The same twisted vortex loop with ${\bf a} \rightarrow
{\bf b} $ and ${\bf b} \rightarrow
-{\bf a}$,  leads to the change of sign of the order parameter, which must be
compensated by a change of its phase $\theta$ by $\pi$. This corresponds to
the $N=1/2$ of circulation quantum. The fractional flux trapped by the loop is
now exactly $\Phi_0/2$, since the parameter $K$ in Eq.(\ref{SuperfluidCurrent})
is exactly zero in nonchiral superconductors.  The same reasoning gives
rise to the $\Phi_0/2$ flux attached to the tricrystal line \cite{Kirtley1996}:
around this line one has ${\bf a} \rightarrow {\bf b} $ and ${\bf b}
\rightarrow
-{\bf a}$. Observation of the fractional flux different from $\Phi_0/2$ would
indicate breaking of the time reversal symmetry
\cite{Sigrist1989,Sigrist1995,VolovikGorkov1984}.

Let us also mention that the vortex sheet with fractional elementary vortices
has been predicted to exist in chiral superconductors \cite{Sigrist1989}, even
before the experimental identification of the vortex sheet in $^3$He-A. The
object which traps vorticity is the domain wall separating domains with the
opposite orientations of the $\hat{\bf l}$-vector
\cite{VolovikGorkov1985}. The kink -- the Bloch line -- in the domain wall is
the vortex with the fractional winding number $N=1/2$. When there are many
fractional vortices (kinks) trapped, they form the vortex sheet, which as
suggested in \cite{Sigrist98} can be responsible for the flux flow dynamics  in
the low-T phase of the heavy fermionic superconductor  UPt$_3$.

In conclusion of this Section let us mention an exotic combined object in
$^3$He-A which still have not been observed -- vortex terminated by hedgehog
\cite{Blaha,VolMinMonopole}.  Four equally ``charged'' half-quantum
vortices, i.e. each with
$N=+1/2$ winding number, can meet each other at one point --
the hedgehog in the $\hat{\bf l}$ field (see Review \cite{Monopoles}). Such
combined object, which reminds the Dirac magnetic monopole with one or several
physical Dirac strings, is called a nexus in relativistic theories
\cite{Cornwall}. In electroweak theory the monopole-antimonopole pair
connected by $Z$ string is called a dumbell \cite{Nambu}.  These objects can
exist in chiral superconductors too, where the  nexus has a magnetic charge
emanating radially from the hedgehog: this charge is compensated by the flux
supplied to the hedgehog by
$N=1/2$ Abrikosov vortices. Nexus provides a natural trap for massive
magnetic monopole -- the `t Hooft-Polyakov
magnetic monopole. If such a monopole enters the chiral superconductor, it is
bound to the hedgehog by two $N=1$ or four $N=1/2$ vortices
\cite{Monopoles}.

\section{VORTICES IN ATOMIC BOSE-CONDENSATES}

 In a vector Bose-condensate with the
hyperfine spin $F=1$  an order
parameter consists of 3 complex amplitudes of  spin
projections
$M=(+1,0,-1)$. They can be organized to form
the complex vector
${\bf a}$:
\begin{equation}
\Psi_\nu=\left(
\matrix{
\Psi_{+1}\cr
\Psi_{0}\cr
\Psi_{-1}\cr
}\right)= \left(
\matrix{
{a_x+ia_y\over \sqrt{2}}\cr
a_z\cr
{a_x-ia_y\over \sqrt{2}}\cr
}\right)
\,\,.
\label{ComponentRepresentation}
\end{equation}
There are two symmetrically distinct phases of the $F=1$ Bose-condensates:
(i) The chiral state, which mimics the orbital part of the $^3$He-A order
parameter in Eq.(\ref{OrderParameter}), ${\bf a}=f(\hat{\bf e}_1 + i\hat {\bf
e}_2)$,  $\hat{\bf l}= \hat{\bf e}_1\times\hat {\bf
e}_2)$, occurs when the scattering length
$a_2$ in the scattering channel of two atoms with the total spin 2 is less than
that with the total spin zero, $a_2<a_0$ \cite{Ho,Ohmi}.
(ii) The nonchiral state, which mimics the spin part of the $^3$He-A order
parameter in Eq.(\ref{OrderParameter}), ${\bf a}=f\hat{\bf d} e^{i\theta}$,
occurs  for $a_2>a_0$.
Each of the two states shares some properties
of superfluid $^3$He-A. In particular, the chiral state (i) has continuous
$4\pi$ vortices \cite{Ho,Ohmi}, which are called skyrmions
there.   An optical method to
create half of the full skyrmion  -- the Mermin-Ho continuous vortices with
$2\pi$ winding -- in the $F=1$ Bose-condensates has been recently discussed in
Ref.
\cite{Marzlin}.
The nonchiral state (ii) may contain $N=1/2$ vortex -- the  combination of
$\pi$-vortex in the phase $\theta$ and
$\pi$-disclination in the vector $\hat{\bf d}$ as in
Eq.(\ref{HalfQuantumVortex}):\cite{LeonhardtVolovik}
\begin{equation}
{\bf a}=f\left(\hat{\bf x} \cos{\phi\over 2} +\hat{\bf y}
\sin{\phi\over 2}\right)e^{i\phi/2}~.
\label{HalfQuantumVortex2}
\end{equation}
In spin projection representation the order parameter far from the
core is
\begin{equation}
\Psi_\nu=fe^{i\phi/2}\left(
\matrix{
e^{i\phi/2}\cr
0\cr
e^{-i\phi/2}\cr
}\right)=f\left(
\matrix{
e^{i\phi}\cr
0\cr
1\cr
}\right)
\,\,.
\label{VortexInOneComponent}
\end{equation}
This means that the $N=1/2$ vortex can be represented as a vortex in the
$M=+1$ component, while the $M=-1$ component is
vortex-free.

The same two cases can occur in a mixture of two condensates. The skyrmion
and $N=1/2$-vortex can be written in the same form:
\begin{equation}
\left(
\matrix{
\Psi_\uparrow\cr
\Psi_\downarrow\cr
}\right)=f\left(
\matrix{ e^{i\phi} \cos {\beta(r)\over 2}
\cr
\sin {\beta(r)\over 2}\cr
}\right)~,~{\hat{\bf l}}=(\sin\beta\cos\phi,-\sin\beta\sin\phi,\cos\beta)
\,\,.
\label{Skyrmion}
\end{equation}
For skyrmion one has $\beta(\infty)=0$ and $\beta(0)=\pi$. This represents
the $N=1$ vortex in  condensate $|\uparrow>$, whose soft core is filled by the
condensate
$|\downarrow>$ as was observed in Ref.\cite{Matthews}.
Note that this is the full skyrmion, i.e. the ${\hat{\bf l}}$ vector sweeps the
whole sphere, however it has $N=1$ winding instead of $N=2$ in a vector
Bose-condensate
\cite{HoTalk}.

The $N=1/2$-vortex is obtained if $\beta(\infty)=\pi/2$:
it is the $N=1$ vortex in the $|\uparrow>$ component, while the
$|\downarrow>$ component is vortex-free.

\section{FERMIONS ON EXOTIC VORTICES}

Interesting phenomena to be observed are related to behavior of fermionic
quasiparticles in  topologically nontrivial environments provided by fractional
vortices and monopoles. In the presence of a monopole the quantum statistics of
particles can change, e.g.  isospin degrees of freedom are trasformed to spin
degrees
\cite{JackiwRebby}.  There are also the fermion zero modes:  bound
states of fermions at monopole or vortex with exactly zero energy.

The low-energy fermions bound to the vortex core play the main role
in the thermodynamics and dynamics of the vortex state in superconductors and
Fermi-superfluids. The spectrum of low-energy bound states in the
core of the $N=\pm 1$  vortex in $s$-wave
superconductor was obtained in Ref.\cite{Caroli}
\begin{equation}
E_n=\omega_0\left(n+{1\over 2}\right).
 \label{Caroli}
\end{equation}
This spectrum is two-fold degenerate due to spin degrees of freedom.
The integral quantum number $n$ is related to the  angular
momentum of the fermions $n=-N L_z$. The level spacing is small compared to the
energy gap of the quasiparticles outside the core, $\omega_0\sim
\Delta^2/E_F\ll\Delta$, and the  discrete nature of the spectrum
becomes revealed only at  $T\sim\omega_0$. There are no
quasiparticle states with energies below $\omega_0/2$. The latter, however,
can appear if the core structure is deformed \cite{Volovik1989}, or in some
regions of the magnetic field due to Zeeman splitting
\cite{MakhlinVolovik}: one
or several energy levels cross zero as a function of momentum
$p_z$, and one obtains the 1D Fermi liquid(s) living in the vortex
core, which can exhibit all the exotic properties of 1D Fermi systems including
Peierls instability \cite{Naculich,MakhlinVolovik} and Luttinger liquid
physics \cite{Senthil2}.

$N= 1$ vortex in $^3$He-A has also
equidistant energy levels but the Maslov index in
Bohr-Sommerfeld quantization is $0$ instead of $1/2$ in Eq.(\ref{Caroli})
\cite{KopninSalomaa}:
\begin{equation}
E_n=\omega_0 n ~.
 \label{CaroliII}
\end{equation}
The spectrum now contains the state with $n=0$, which has
exactly zero energy.  This $E=0$ level is doubly degenerate due to spin.
Such fermion zero mode is robust to any deformations, which preserve the
spin degeneracy \cite{Volovik1999}.

The most exotic situation will occur for the $N=1/2$
vortex in layered chiral superconductors. As we discussed above, the $N=1/2$
vortex is equivalent to the $N=1$ vortex in  only one spin component.  That is
why each layer of superconductor contains only
one energy level with $E=0$.  Since the $E=0$ level can be
either filled or empty, there is a fractional  entropy  $(1/2) ln~ 2$ per layer
per vortex. The factor  (1/2) appears because in superconductors the
particle excitation coincides with its antiparticle (hole), i.e. the
quasiparticle is a Majorana fermion \cite{Read}. Majorana fermions at $E=0$
level lead to the non-Abelian statistics of $N=1/2$ vortices:
the interchange of two point vortices becomes identical operation (up to an
overall phase) only on being repeated four times \cite{Ivanov}.  This can be
used for quantum computing \cite{Kitaev}.

Also the spin of the vortex in a chiral superconductor can be fractional,
as well
as the electric charge per layer per vortex \cite{Goryo}, but this is still not
conclusive. The problem with the fractional charge, spin and statistics,
related
to the topological defects in chiral superconductors is still open. This is
related also to the problem of the quantization of physical parameters in 2D
systems, such as Hall conductivity and spin Hall conductivity
\cite{Ishikawa,VolovikYakovenko,Yakovenko,Senthill,Read}.

%
%

\section*{ACKNOWLEDGMENTS}

I thank B. Anderson for illuminating discussions. This work  was supported in
part by  Russian Foundation for Fundamental Research and ESF.

\end{document}